\documentclass{aastex}          %% The manuscript based on AASTeX v5.x
\usepackage{spr-astr-addons}    %% mimicing ASTR journal style
\begin{document}
%% Article title
%
\title{High-contrast imaging constraints on gas giant planet formation -- The Herbig Ae/Be star opportunity}

%% Running heads
\shorttitle{Gas giant planet formation around Herbig Ae/Be stars}
\shortauthors{Sascha P. Quanz}

%% Author and Affilations
\author{Sascha P. Quanz\altaffilmark{1}} 
%\and 
%\author{\altaffilmark{}}
%\affil{}
%\email{} %% non-output

%% Alternate Affilations
\altaffiltext{1}{Institute for Astronomy, ETH Zurich, Wolfgang-Pauli-Strasse 27, 8093 Zurich, Switzerland}
%\altaffiltext{2}{}
%\altaffiltext{3}{}

%% Abstract
\begin{abstract}
Planet formation studies are often focused on solar-type stars, implicitly considering our Sun as reference point. This approach overlooks, however, that Herbig Ae/Be stars are in some sense much better targets to study planet formation processes empirically, with their disks generally being larger, brighter and simply easier to observe across a large wavelength range. In addition, massive gas giant planets have been found on wide orbits around early type stars, triggering the question if these objects did indeed form there and, if so, by what process. In the following I briefly review what we currently know about the occurrence rate of planets around intermediate mass stars, before discussing recent results from Herbig Ae/Be stars in the context of planet formation. The main emphasis is put on spatially resolved polarized light images of potentially planet forming disks and how these images - in combination with other data - can be used to empirically constrain (parts of) the planet formation process. Of particular interest are two objects, HD100546 and HD169142, where, in addition to intriguing morphological structures in the disks, direct observational evidence for (very) young planets has been reported. I conclude with an outlook, what further progress we can expect in the very near future with the next generation of high-contrast imagers at 8-m class telescopes and their synergies with ALMA. 
\end{abstract}

%% Keywords

%%  Please use labels (\label, \ref) for section, figure, table, 
%%  equation  reference. Use \cite for bibliography references.
%
\section{Introduction}\label{s:intro}
Where, when and how do gas giant planets form? ---  Two main theories of gas giant planet formation - the core accretion (CA) paradigm, based on the collisional growths of dust particles, and the gravitational instability (GI) theory - provide a physical foundation to address these questions from a theoretical perspective. In addition, astronomical observations start providing a wealth of empirical data to constrain those theories at least at the earliest and the final stages of planet formation, by revealing ever more details about the physical and chemical conditions in protoplanetary disks, where planet formation is thought to occur, and by studying the occurrence rate of planets, i.e. the outcome of the planet formation process. However, the formation process itself is still largely unconstrained as we lack observational data. As a result there are large uncertainties in the luminosity evolution of gas giant planets over the first few hundred million years mainly because the initial entropy of the objects, which is set by the physics of the gas accretion process during formation, is not known. This in turn leads to a wide spread in predicted magnitudes for these objects for a given age and mass \citep[``hot star" vs. ``cold start" models; e.g.,][]{Marley et al.(2007), Spiegel & Burrows(2012)}.

Quite naturally, in the context of planet formation and exoplanet studies, focus is put on Sun-like stars in order to put our own Solar System in context. However, in order to understand gas giant planet formation and to derive empirical constraints based on astronomical observations, intermediate mass stars, and their young counterparts the Herbig Ae/Be stars, offer unique opportunities as I will discuss in the following. 

%\subsection{}%\label{ss:?}
%\subsubsection{}%\label{sss:?}
\begin{table*}[t]
\caption{Directly imaged gas giant planets around intermediate mass stars. \label{table1}}
\begin{center}

\begin{tabular}{lllll}
\tableline
Star & Spectral Type & Planet & Separation & Detection reference\\
\tableline
$\beta$ Pictoris & A6V & $\beta$ Pic b & $\sim$8--9 au\tablenotemark{a} & \citet{Lagrange et al.(2009),Lagrange et al.(2010)}\\
HR8799 & F0V mA4 Lam Boo & HR8799 b & $\sim$14 au & \citet{Marois et al.(2008)}\\
  &   & HR8799 c & $\sim$24 au & \citet{Marois et al.(2008)}\\
  &   & HR8799 d & $\sim$38 au & \citet{Marois et al.(2008)}\\
  &   & HR8799 e & $\sim$68 au & \citet{Marois et al.(2010)}\\
$\kappa$ Andromedae & B9 IV &  $\kappa$ And b & $\sim$55 au & \citet{Carson et al.(2013)}\\
HD95086 & A8V & HD95086 b & $\sim$56 au & \citet{Rameau et al.(2013a),Rameau et al.(2013b)}\\
\tableline
\tablenotetext{a}{Estimated semi-major axis of the planet's orbit \citep{Chauvin et 
al.(2012)}.}
\end{tabular}
\end{center}

\end{table*}

\section{The exoplanet population around intermediate mass stars}\label{s:exoplanets}
A good starting point for discussing gas giant planet formation around intermediate mass stars is to look at the population of exoplanets that has been discovered around these stars and how it might differ from the population around lower-mass, i.e., solar-type, stars.  Unfortunately, while the \emph{Kepler} mission \citep{Borucki et al.(2010)} has revolutionized our understanding of the exoplanet population around M, K, G and F type dwarfs, not much focus was put on more massive stars. It seems as if the occurrence rate of ``large" planets (with radii $>$4 R$_\earth$) does not depend on stellar type and is on the order of a few percent \citep[e.g.,][]{Fressin et al.(2013),Howard et al.(2012)}. Furthermore, the period distribution of these planets, expressed in logarithmic terms, seems to increase out to $\sim$100 days, but then it flattens off \citep{Fressin et al.(2013)}. However, these analyses do not not explicitly include A type stars (or earlier) and they are confined to orbital periods $\lesssim$400 days. Turning to results from radial velocity (RV) searches it has become clear that the  occurrence rate of gas giant planets increases with stellar mass for stars more massive than the Sun \citep{Johnson et al.(2007a),Johnson et al.(2007b),Johnson et al.(2010),Bowler et al.(2010)} and $\sim$15\% of stars with masses of $\sim$2 M$_\sun$ have a giant planet \citep{Johnson et al.(2010)}. Recently, however, \citet{Reffert et al.(2015)} found that the occurrence rate seems to drop rapidly again for stars more massive than 2.5--3.0 M$_\sun$. Concerning the planets' period distribution an interesting difference between more massive and Sun-like stars is the apparent paucity of planets between 0.1 and 0.6 au around stars with masses $>$1.5 M$_\sun$ \citep{Johnson et al.(2011)}. The general trend, however, indicates that the number of gas giant planets increases with increasing orbital period \citep[e.g.,][]{Mayor et al.(2011)} and that more massive stars harbor more massive planets on longer period orbits \citep{Johnson et al.(2010)}. Because up to now RV surveys are limited to planet detections in the inner few au, it is not clear yet at what separations the giant planets occurrence rate may reach a maximum. Extrapolating the RV results to larger orbital separations led to the suggestion that intermediate mass stars are the best targets to search for gas giant planets via direct imaging \citep[e.g.,][]{Crepp & Johnson(2011)}. And, indeed, the majority of the directly imaged planets\footnote{The term `planet' is used for objects with an estimated mass ratio $<$0.02 and a separation $<$100 au between the object and its parent star  \citep[cf.][]{Pepe et al.(2014)}.} were found around intermediate mass stars (see, Table~\ref{table1}). However, most large-scale direct imaging surveys searching for giant exoplanets around dozens of nearby, young, intermediate mass stars yielded null-results \citep[e.g.,][]{Janson et al.(2011),Vigan et al.(2012),Nielsen et al.(2013),Rameau et al.(2013c)}. These surveys were carried out in the near-infrared H or K band or in the thermal infrared L band, and the authors used Angular Differential Imaging \citep[ADI;][]{Marois et al.(2006)} to enhance the contrast performance and optimize the data for the detection of faint, nearby companions\footnote{In addition to ADI, \citet{Nielsen et al.(2013)} also used Angular and Spectral Differential Imaging (ASDI) where objects are imaged simultaneously in two narrow band filters, one centered on the 1.652 $\mu$m methane band and the other on the nearby continuum. By subtracting one filter from the other additional sensitivity can be achieved for low-mass objects with a (strongly) methanated atmosphere as the stellar PSF and speckle noise can be accurately removed.}. These null-results demonstrate that systems like the 4-planet HR8799 system are certainly the exception and not the rule. Based on the achieved detection limits, these surveys allow us to put statistical constraints on the occurrence rate of gas giant planets at wide orbital separations. \citet{Nielsen et al.(2013)}  found that $<$20\% of $\sim$2 M$_\sun$ stars have companions with masses $\gtrsim$4 M$_{\rm Jupiter}$  between 59 and 460 au (95\% confidence) and comparable results were presented by \citet{Vigan et al.(2012)}. For smaller separations from the host stars or for planets with lower masses the existing imaging surveys did not yet provide any strong constraints on the planet occurrence rate. It should be noted that in order to convert detection limits into companion mass limits most studies so far used ``hot-start" evolutionary models \citep[e.g.,][]{Baraffe et al.(2003)} that, for a given planet mass and for the young ages of typical target stars, predict brighter planets than ``cold-start" models \citep[e.g.,][]{Marley et al.(2007)}.

In summary this means that (1) there are a few intermediate mass stars, where gas giant planets have been detected on orbits with semi-major axis $a \gtrsim10$ au, (2) such systems are rare, and (3) somewhere between a few and a few tens of au the peak for the occurrence rate of gas giant planets around intermediate mass stars can be expected. 

%% Figure 

%A key caveat here is that in order to determine the detection limits of direct imaging surveys, model predictions for the planets' brightness at a given age and for a given mass are needed and as mentioned above the initial conditions for these models are basically unconstrained. 

%
\begin{figure}[h!]
\centering
\includegraphics[height=4.1cm]{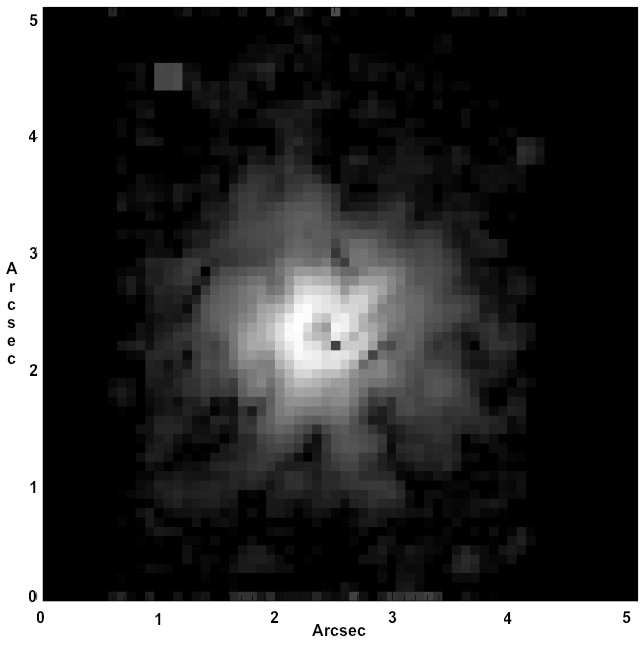}
\includegraphics[height=4.0cm]{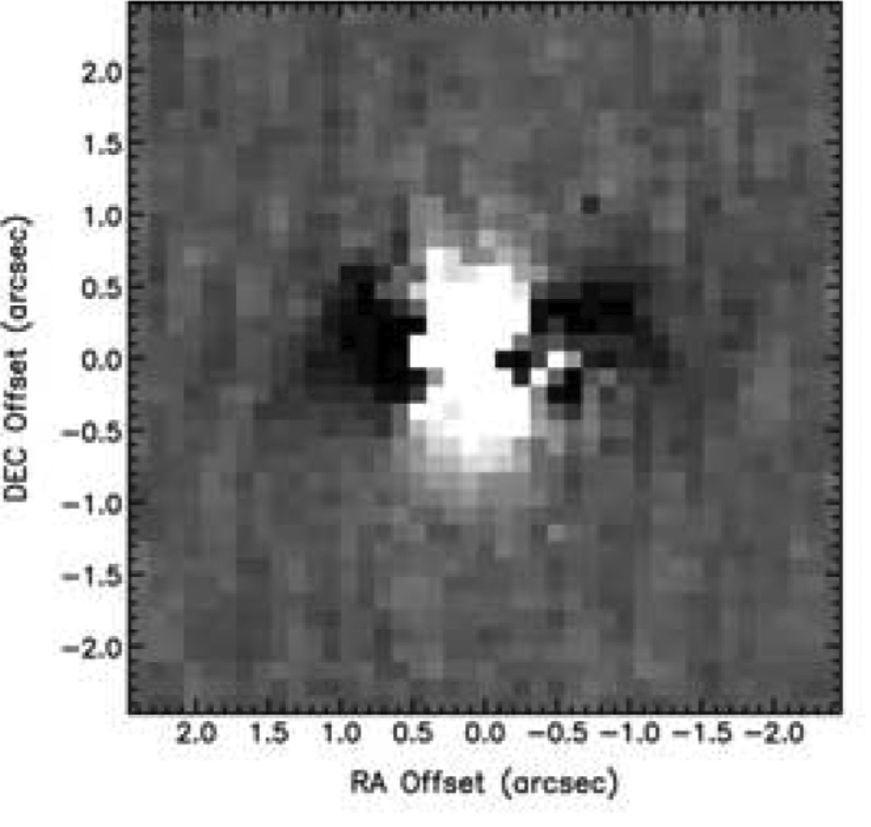}
\includegraphics[height=3.7cm]{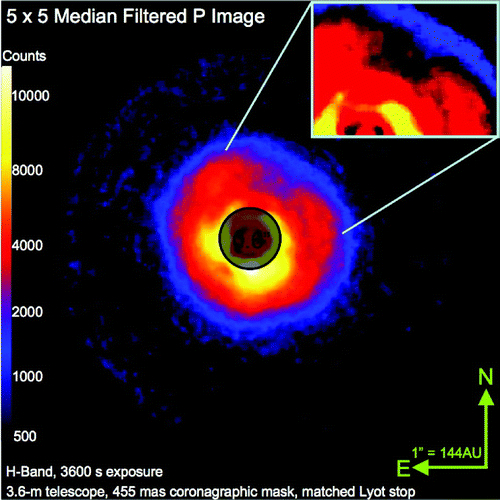}
\includegraphics[height=4.0cm]{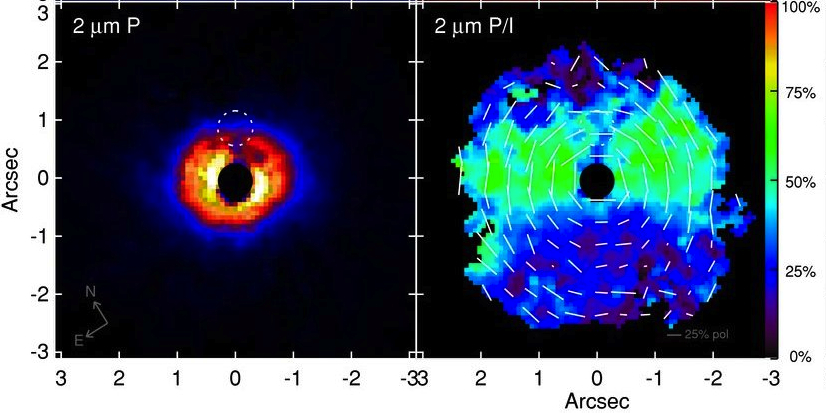}
\caption{Earlier near-infrared, ground- and space-based PDI studies of Herbig Ae/Be disks. Top row: HD169142 in total polarized flux $P$ \citep[Image credit:][]{Kuhn et al.(2001)} and Stokes $Q$ \citep[Image credit:][reprinted by permission of Oxford University Press]{Hales et al.(2006)}. Middle panel: AB Aurigae in polarized flux $P$ \citep[Image credit:][]{Oppenheimer et al.(2008)}. Bottom row: Again AB Aurigae in polarized flux $P$ and polarization fraction $P/I$ observed with the \emph{NICMOS} instrument onboard the \emph{Hubble Space Telescope (HST)} \citep[Image credit:][]{Perrin et al.(2009)}. While the overall extent of the disks can be easily derived, the sensitivity and/or spatial resolution to resolve finer disk structures is limited. Unless otherwise specified all images \copyright AAS. Reprinted with permission.} %% no full stop at the end of caption
 \label{fig:history}
 \end{figure}
 
\section{Challenges for planet formation theories}\label{s:theory}
Comparing the predictions of the two main theories of gas giant planet formation mentioned in the Introduction to the directly imaged gas giant planets and the results from large scale imaging surveys, it shows that theory and observations are not easy to reconcile\footnote{Even if not the main focus here, it should be mentioned that not only theories of planet formation, but also theories of exoplanet atmospheres are challenged by some of the directly imaged planets as the observed and predicted colors in the 1 -- 5 $\mu$m regime are quite discrepant \citep[e.g.,][]{Skemer et al.(2012)}.}. In the CA picture the time required to assemble a planetary core of several Earth masses - onto which a gaseous envelope is then accreted - exceeds the lifetime of the gas contained within the circumstellar disk if the planet forms at orbital separations beyond $\sim$15 au \citep[e.g.][]{Ida & Lin(2004),Kennedy & Kenyon(2008)}. However, even if GI in general prefers the formation of massive planets on wide orbits, it is not obvious that the HR8799 planetary systems can be formed through this mechanism \citep[e.g.,][]{Kratter et al.(2010)}. Furthermore, \citet{Janson et al.(2011)} found that coupling their null-result from a deep exoplanet imaging survey around nearby massive stars with target specific planet formation simulations suggests that GI cannot be the main formation mechanism for gas giant planets: $<$30\% of stars form and retain GI companions ($<$100 M$_{\rm Jupiter}$) within 300 au with 99\% confidence.

A possible way to circumvent the problem of the CA theory to form planets at large orbital separations is to speed up the assembly of planetary cores at these locations. This might be achieved via the so-called 'pebble-accretion' mechanism that was proposed by \citet{Lambrechts & Johansen(2012)}. Here, cm-sizes dust particles that are only loosely coupled to the gas in the circumstellar disk can be accreted very efficiently onto a planetary embryo and reduce the formation time of a planetary core by several orders of magnitudes.

\section{High-spatial resolution images of circumstellar disks using polarimetric differential imaging}\label{s:pdi}
To make progress from the observational side and to better constrain gas giant planet formation models one needs to be able to spatially resolve those regions in circumstellar disks where planet formation is thought to occur. For intermediate mass stars this means separations from a few out to several tens of au. At the typical distances of young stars of  (100--200 pc) this translates into a spatial resolution and inner working angle (IWA) requirement of $\lesssim$0.1$''$. A very powerful technique to probe these disk regions at these scales is polarimetric differential imaging (PDI). The basic principle is the following: To zeroth order light from the central star is unpolarized, while photons scattered off the dust grains on the surface of circumstellar disks have a resulting linear polarization. This means that by using a double-beam imaging camera, observing an object simultaneously in two channels in polarized light where the polarization direction is rotated by 90$^\circ$ between the two channels, and then subtracting the two resulting images from each other, the central star should almost perfectly cancel out, while the polarized light from a circumstellar disk might result in a detectable signal. By collecting data for different position angles for the polarization direction, one can construct the Stokes vectors $Q$ and $U$, which can then be combined in different ways to yield the total polarized flux $P$ of the circumstellar disk \citep[e.g.,][]{Tinbergen(1996),Schmid et al.(2006),Hinkley et al.(2009),Quanz et al.(2011)}.

\begin{table*}[t!]
\caption{Herbig Ae/Be stars and other young intermediate mass stars observed with PDI on 8-m telescopes. \label{table2}}
\begin{center}

\begin{tabular}{lllll}
\tableline
Object & Instrument & Filter & Radial extent of & References\\
  &  &    &  PDI detection &  \\
\tableline
AB Aur & Subaru/HiCIAO & H & $\sim$0.15$''$--3.85$''$ ($\sim$22--553 au) & \citet{Hashimoto et al.(2011)}\\ 
HD97048 & VLT/NACO & H,K$_s$  &  $\sim$0.1$''$--1.0$''$ ($\sim$16--160 au) & \citet{Quanz et al.(2012)}\\
HD100546 & VLT/NACO & H,K$_s$ & $\sim$0.1$''$--1.4$''$ ($\sim$10--140 au) &\citet{Quanz et al.(2011)}  \\
   & VLT/NACO & H,K$_s$ &   $\sim$0.1$''$--1.5$''$\tablenotemark{a} ($\sim$10--150 au)& \citet{Avenhaus et al.(2014b)}  \\
   & VLT/NACO & L$'$ &    $\sim$0.1$''$--0.5$''$ ($\sim$10--50 au) & \citet{Avenhaus et al.(2014b)}  \\

  SAO206462\tablenotemark{b} & Subaru/HiCIAO & H & $\sim$0.2$''$--1.0$''$ ($\sim$28--140 au) & \citet{Muto et al.(2012)}\\
 & VLT/NACO & H,K$_s$ & $\sim$0.1$''$--0.9$''$ ($\sim$14--125 au)  & \citet{Garufi et al.(2013)}\\ 
  HD141569A & VLT/NACO & H & not detected\tablenotemark{c}  &\citet{Garufi et al.(2014)}\\ 
 HD142527 & VLT/NACO & H,K$_s$ &  $\sim$0.3$''$--1.8$''$ ($\sim$45--270 au)  & \citet{Canovas et al.(2013)}\\
 		& VLT/NACO & H,K$_s$ &  $\sim$0.1$''$--2.5$''$ ($\sim$15--360 au) & \citet{Avenhaus et al.(2014a)} \\ 
		& Gemini/GPI & Y &   $\sim$0.06$''$--1.5$''$ ($\sim$9--225 au)  & \citet{Rodigas et al.(2014)}\\
HD150193A & VLT/NACO & H,K$_s$ & not detected\tablenotemark{c}  &\citet{Garufi et al.(2014)}\\		
HD163296 & VLT/NACO & H,K$_s$ & $\sim$0.4$''$--1.0$''$ ($\sim$50--122 au) & \citet{Garufi et al.(2014)}\\		
 HD169142 &VLT/NACO &  H & $\sim$0.1$''$--1.7$''$ ($\sim$15--250 au) & \citet{Quanz et al.(2013b)}\\
  & Subaru/HiCIAO & H & $\sim$0.2$''$--1.1$''$ ($\sim$29--160 au) & \citet{Momose et al.(2013)}\\
  MWC480 & Subaru/HiCIAO &  H & $\sim$0.2$''$--1.0$''$ ($\sim$28--137 au) & \citet{Kusakabe et al.(2012)}\\
  MWC758 & Subaru/HiCIAO & H &  $\sim$0.2$''$--0.8$''$ ($\sim$56--223 au / $\sim$40--160 au)\tablenotemark{d}  & \citet{Grady et al.(2013)}\\
  Oph IRS48\tablenotemark{e} &  Subaru/HiCIAO & H, K$_s$& $\sim$0.2$''$--1.3$''$ ($\sim$24--157 au) & \citet{Follette et al.(2015)}\\
  SR21\tablenotemark{f} & Subaru/HiCIAO & H & $\sim$0.1$''$--0.6$''$ ($\sim$12--75 au) & \citet{Follette et al.(2013)}\\  
 \tableline
\tablenotetext{a}{The disk is even detected beyond 1.5$''$, but the analyses are focused on the range given here.}
\tablenotetext{b}{Also known as HD135344 B. It has a spectral type of F4, but is often discussed in the context of Herbig Ae/Be stars.}
\tablenotetext{c}{3$\sigma$ upper limits on the polarized flux are given between $\sim$15--150 au ($\sim$0.15$''$--1.5$''$) for HD141569A and $\sim$15--240 au ($\sim$0.1$''$--1.6$''$) for HD150193A. }
\tablenotetext{d}{The distance to the object is not well known and \citet{Grady et al.(2013)} considered both 160 pc and 200 pc.}
\tablenotetext{e}{Also cataloged as WLY 2-48; not a bona fide Herbig Ae/Be star, but a young A-type star with significant foreground extinction.}
\tablenotetext{f}{A young, intermediate mass star ($\sim$2.5 M$_\sun$) in Ophiuchus, but with a spectral type of G3 not a bona fide Herbig Ae/Be star.}

\end{tabular}
\end{center}

\end{table*}

PDI is not a new technique. In fact, it has been around for more than a decade and has been used from the ground and from space to image the dusty environment of stars including circumstellar disks and debris disks. In Figure~\ref{fig:history} a few selected examples of earlier studies of Herbig Ae/Be stars using the PDI technique are shown. While spatially resolving a circumstellar disk was certainly a big success on its own and laid the ground for everything that followed, these earlier studies were still limited in sensitivity and/or spatial resolution. In the past few years, however, PDI became one of the leading techniques to image circumstellar disks from the ground. One reason was that high-resolution, AO-assisted, near-infrared cameras on 8-m telescopes were equipped with PDI capabilities. In addition to this, new ways were found to correct for quite severe instrumental/telescope polarization effects, which all of these systems suffered from \citep{Joos et al.(2008),Witzel et al.(2011),Quanz et al.(2011), Avenhaus et al.(2014a)}. While the absolute polarization accuracy of these instruments may still be an order of magnitude away from what high-precision polarimetrists would deem appropriate, the calibration strategies applied these days are good enough to probe circumstellar disks as close as $\sim$0.1$''$ to the star and reveal a tremendous amount of sub-structure in some of these disks in the inner few tenths of an arcsecond.  

\begin{figure*}[t!]
\centering
\includegraphics[height=5.4cm]{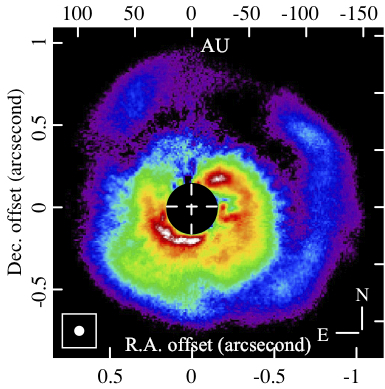}
\includegraphics[height=5.4cm]{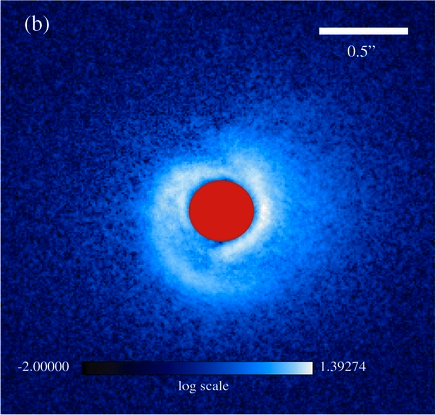}
\includegraphics[height=5.4cm]{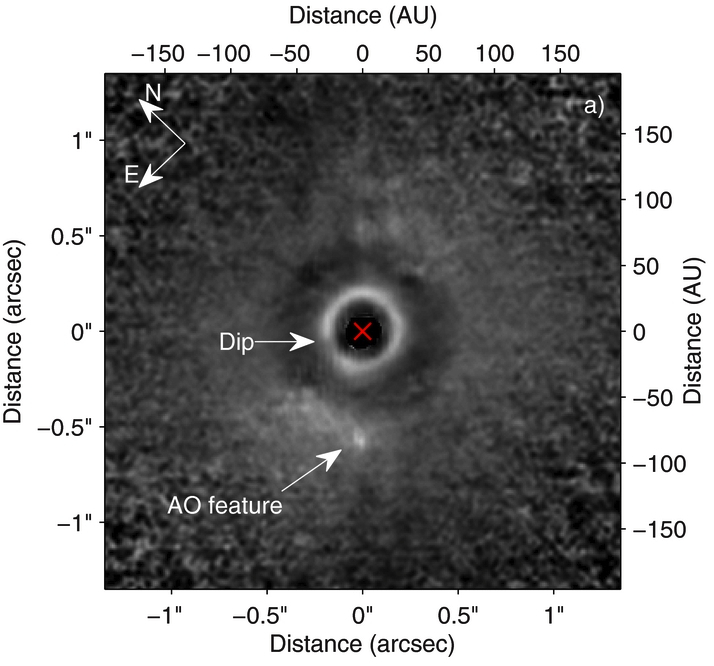}
\includegraphics[height=5.0cm]{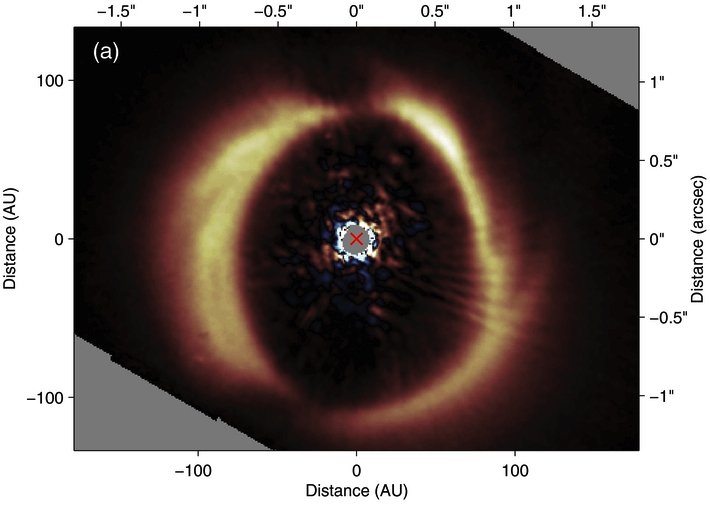}
\includegraphics[height=5.0cm]{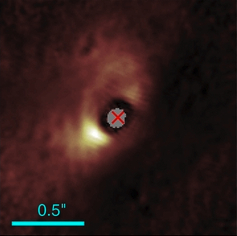}
\includegraphics[height=5.0cm]{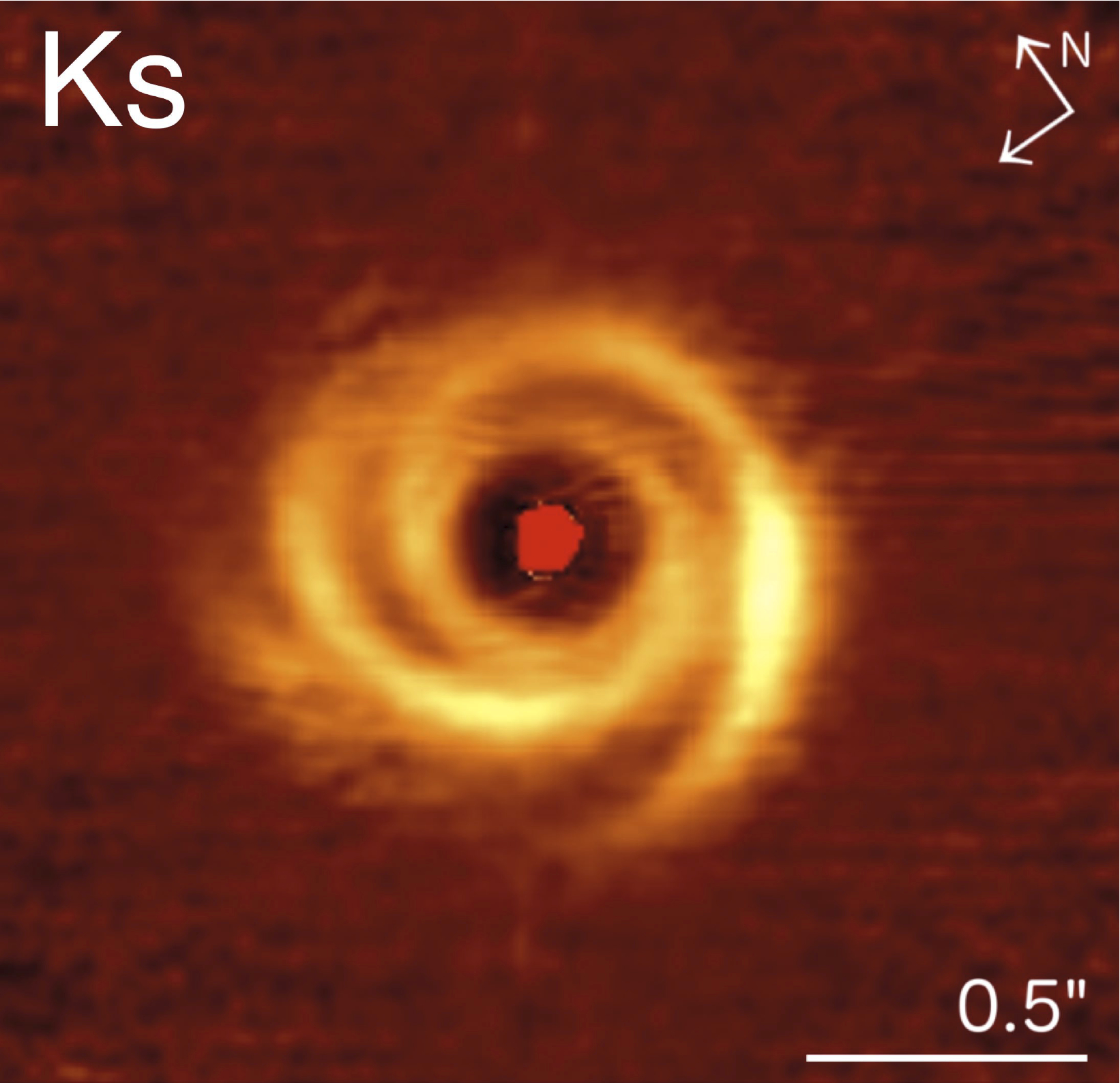}
\caption{Selection of disks around Herbig Ae/Be stars observed with PDI on 8-m class telescopes with AO-assisted, high-resolution, near-infrared cameras. Top row: AB Aurigae \citep[Image credit:][]{Hashimoto et al.(2011)}, MWC758 \citep[Image credit:][]{Grady et al.(2013)}, and HD169142 \citep[Image credit:][]{Quanz et al.(2013b)}. Bottom row: HD142527 \citep[Image credit:][]{Avenhaus et al.(2014a)}, HD100546 \citep[Image credit:][]{Avenhaus et al.(2014b)}, and SAO206462 \citep[Image credit:][]{Garufi et al.(2013)}. Spiral arms, arcs, cavities, gaps and holes are commonly detected in these PDI images illustrating the large amount of sub-structure in the inner few tens of au around these objects. The image of SAO206462 is reproduced with permission \copyright ESO; all other images \copyright AAS. Reproduced with permission.
 } %% no full stop at the end of caption
 \label{fig:PDI}
 \end{figure*}

Table~\ref{table2} summarizes recent state-of-the-art PDI observations of Herbig Ae/Be stars and in Figure~\ref{fig:PDI} a few selected examples of the resulting disk images are shown. Spiral arms, inner cavities, gaps and holes appear to be common structures and yet the variety of morphologies is quite surprising. A comparison of the images of HD169142 and AB Aurigae shown in Figure~\ref{fig:history} and Figure~\ref{fig:PDI} nicely demonstrates the level of improvement in terms of data acquisition and analysis that occurred over the last years. 

It needs to be emphasized that PDI probes the surface layer of a circumstellar disk, meaning that the structures that are seen can not be readily interpreted as surface mass density variations. Also, the resulting polarization signal depends on the scattering and polarization properties of the dust grains each of which depends on (a) the scattering angle (influenced by disk inclination and disk flaring) and (b) dust grain properties such as size, composition and structure (e.g., compact grains vs. fluffy aggregates) \citep[e.g.,][]{Pinte et 
al.(2008),Perrin et al.(2009),Min et al.(2012)}. Seeing a gap-like structure or a region of reduced polarized flux can hence have different underlying reasons and care must be taken with the interpretation \citep[e.g.,][]{Garufi et al.(2014)}. For a rough approximation one can assume that PDI probes dust grains on the disk surface layer that have effective sizes comparable to the observing wavelength.

\section{Potentially planet forming Herbig Ae/Be stars: 4 case studies}\label{s:case_studies}
Among the objects listed in Table~\ref{table2} there are four that warrant special attention from a planet formation perspective as their PDI images, in combination with other datasets, provide interesting empirical constraints. These four sources are discussed in the following. An important common feature of all of these systems is that the central stars are still actively accreting material and that they harbor a small inner disk in the inner few au that is undetected in the PDI images, but contributes to the observed excess emission at NIR wavelengths. 

\subsection{HD142527}
HD142527 is an extensively studied Herbig Ae/Be star and was observed by various groups in PDI mode (see, Table~\ref{table2} and Figure~\ref{fig:PDI}). The most striking feature of this object is the complex spiral arm structure and the huge disk gap that is clearly detected in the PDI images and that stretches from a few out to more than 100 au in radius. This disk gap is also apparent at longer wavelengths, normally probing larger dust grains residing in the mid-plane of circumstellar disks, but not in molecular line emission, showing that gas in present in those dust depleted regions of the disk \citep{Casassus et al.(2013)}. From other imaging data and SED modeling this gap has been suggested before \citep[e.g.,][]{Fukagawa et al.(2006),Verhoeff et al.(2011),Rameau et al.(2012)} and it has been speculated to what extent planets might be responsible for clearing this gap and creating the observed complex and asymmetrical dust morphology. In particular, there is a significant azimuthal asymmetry in the mm-sized dust grains, which could be explained via pressure supported dust traps induced by orbiting companions \citep{Casassus et al.(2013)}. Interestingly, sparse aperture masking observations \citep{Biller et al.(2012)} revealed a relatively bright companion candidate at a separation of only $\sim$88 mas ($\sim$13 au), which was subsequently confirmed by high-contrast AO-imaging at optical wavelength to be an accreting $\sim$0.25 M$_\sun$ stellar companion \citep{Close et al.(2014)}. This stellar companion was also detected in Y band total intensity data by \citet{Rodigas et al.(2014)} who furthermore found a point-source in polarized flux in the immediate vicinity of the object. The physical link between the stellar companion and the slightly offset ($\sim$2.7 au) polarized emission is still unclear and whether or not the polarized emission is related to a circumsecondary dust disk or an accretion flow from the main disk feeding the companion remains to be investigated. Also the connection between the companion and the large scale structure of the main disk requires further analyses. Unless the companion is on an extremely eccentric orbit additional, low-mass companions might be required to explain the very large disk gap \citep[e.g.,][]{Avenhaus et al.(2014a), Dong et al.(2014), Zhu et al.(2011)}.  However, other high-contrast imaging searches for planets yielded null-results so far \citep{Rameau et al.(2012), Casassus et al.(2013)} even though companions with masses $\gtrsim$15 M$_{\rm{Jupiter}}$ ($\gtrsim$9 M$_{\rm{Jupiter}}$) had a detection probability of $\gtrsim$50\% for separations $\gtrsim$70 au ($\gtrsim$100 au) from the star \citep{Rameau et al.(2012)}.

%Current detection limits are limits in terms of detectable mass were moderate as  

\subsection{SAO206462 (HD135344B)}
A well defined 2-armed spiral structure is the main morphological feature of SAO206462 in the PDI images \citep[][see also Figure~\ref{fig:PDI}]{Muto et al.(2012), Garufi et al.(2013)}. \citet{Muto et al.(2012)} used spiral density wave calculations to determine where in the disk two unseen planets driving the spiral arms could be located. In addition, from an estimate for the amplitude of the surface density perturbation and the apparent non-existence of a disk gap the authors concluded that the planets' masses might be around 0.5 M$_{\rm{Jupiter}}$. In addition to the spiral arm structure, the PDI images of \citet{Garufi et al.(2013)} revealed a disk cavity with a radius of $\sim$28 au. This cavity size is significantly smaller than the one detected earlier at sub-mm wavelengths, which was $\sim$39--50 au \citep{Brown et al.(2009),Andrews et al.(2011)}. First results from ALMA \citep{Perez et al.(2014)} yielded a radius of $\sim$45 au for the cavity and the data suggested a more complex morphology also at sub-mm wavelengths as a model consisting of a ring and a vortex-like structure yielded a better fit to the data than a ring model alone (see also, Fig~\ref{fig:SAO}). Cavity sizes that change with observing wavelengths -- appearing smaller at shorter wavelengths probing smaller dust grains and larger at longer wavelengths -- can be explained via planet-disk interactions and dust filtering \citep[e.g.,][]{Dong et al.(2012),Pinilla et al.(2012),de Juan Ovelar et al.(2013)}. In this case, a massive gas giant planet would be expected to reside within 15--20 au from the star \citep{Garufi et al.(2013)}. To date, no high-contrast imaging data searching for the suspected planet(s) have been published and whether or not a single companion can explain both the spiral arm structure and the different cavity sizes remains to be seen.

\subsection{HD169142}
In the context of planet formation the morphology of the disk surrounding HD169142, as seen in PDI images, is probably very close to a textbook example: The disk is seen almost face on and in polarized light an inner cavity is detected out to $\lesssim$20 au, followed by a bright ring with maximum brightness at $\sim$25 au, followed by an annular gap between $\sim$40 and 70 au, and then a smooth disk out to $\sim$250 au \citep[][see also Figure~\ref{fig:PDI}]{Quanz et al.(2013b),Momose et al.(2013)}. This double-gap structure with a bright ring in between might be indicative of 2 orbital regions where planet formation recently occurred and the companions carved out a significant fraction of the disk material \citep[e.g.,][]{Pinilla et al.(2015),Meru et al.(2014)}. Observations at cm-wavelengths with the EVLA confirmed the general disk structure also for larger dust grains expected to be located in the disk mid-plane, and these data also suggest a localized overdensity of  dust particles south of the central star and right in the middle of the annular gap at $\sim$50 au \citep{Osorio et al.(2014)}. These results need further confirmation, but if correct they may suggest the presence of a dusty circumplanetary disk surrounding a young, still forming gas giant planet orbiting within the annular gap. Follow-up, high-contrast ADI observations in the L$'$ filter ($\lambda_{\rm cen}=3.8\,\mu$m), initially intended to confirm this object at shorter wavelengths, did not reveal any source at this location, but constrained the effective temperature and emitting area of this candidate object \citep{Reggiani et al.(2014)}. More importantly, however, these observations yielded another companion candidate located inside the inner cavity close to the bright ring \citep[$\sim$23$\pm$5 au;][]{Reggiani et al.(2014)}. This object was simultaneously also detected by \citet{Biller et al.(2014)} (see also Fig. 3). Both \citet{Biller et al.(2014)} and \citet{Reggiani et al.(2014)} tried to confirm this companion candidate at wavelengths shortwards of 3 $\mu$m, but failed. This is interesting because if the observed L$'$ brightness came solely from the photosphere of a young companion, then evolutionary and atmospheric models for young gas giant planets predict near-infrared magnitudes for this object that are above the derived detection limits from \citet{Reggiani et al.(2014)} and \citet{Biller et al.(2014)}. Whether or not this source is some locally heated patch of the circumstellar disk \citep{Biller et al.(2014)} or a true companion possibly surrounded by a circumplanetary disk \citep{Reggiani et al.(2014)} remains to be seen. In the latter case, the observed L$'$ brightness would be a combination of the fluxes coming from the photosphere of the young planet and from its surrounding circumplanetary disk. While model calculations predicting the spectral energy distributions of circumplanetary disks are still scarce, there are first hints that such disks might be very bright between 3 -- 10 $\mu$m, which could explain why this object is detected at L$'$ but not at shorter wavelengths \citep{Zhu(2015), Eisner(2015)}. 

\begin{figure}[t!]
\centering
\includegraphics[height=4.1cm]{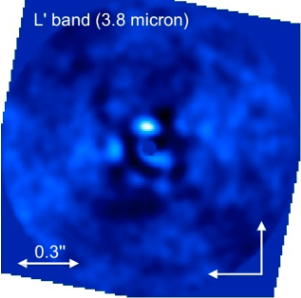}
\includegraphics[height=4cm]{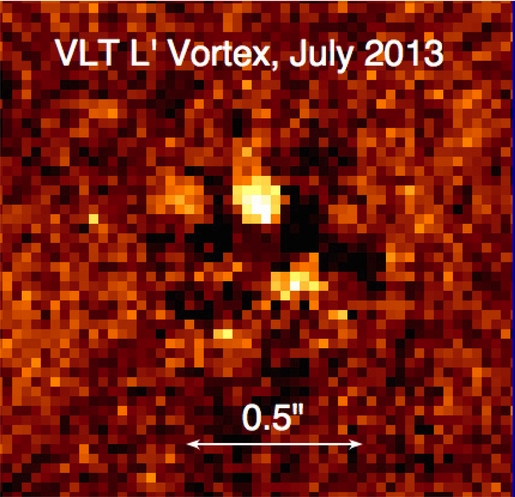}
\caption{The companion candidate around HD169142 detected in high-contrast L$'$ observations \citep[Image credit:][]{Reggiani et al.(2014),Biller et al.(2014)}. \copyright AAS. Reproduced with permission.} %% no full stop at the end of caption
 \label{fig:HD169142b}
 \end{figure}
 
 \begin{figure}[b]
\centering
\includegraphics[height=4.1cm]{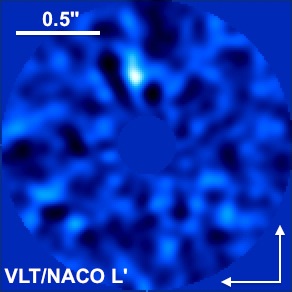}
\includegraphics[height=4.1cm]{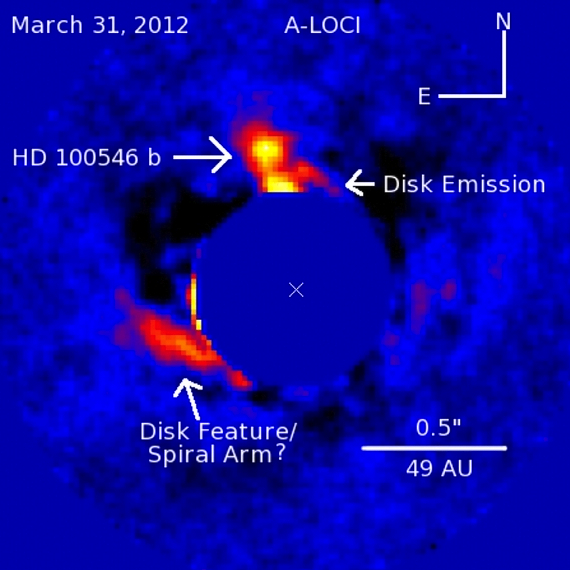}
\caption{The companion candidate around HD100546 detected in high-contrast L$'$ observations \citep[Image credit:][]{Quanz et al.(2013a),Currie et al.(2014)}. \copyright AAS. Reproduced with permission.
} %% no full stop at the end of caption
 \label{fig:HD100546b}
 \end{figure}
 
\subsection{HD100546}
HD100546 is another example of an intensively studied Herbig Ae/Be star and it is impossible to list and cite all relevant publications on this source from the last 10 years. While earlier scattered light observations of this object -- from ground and space -- uncovered the large spatial extent of dusty material around the star \citep[][]{Pantin et al.(2000),Augereau et al.(2001),Grady et al.(2001),Ardila et al.(2007)}, the first PDI observations revealed the inner disk regions and probed the disk surface layer down to $\sim$10 au \citep[][]{Quanz et al.(2011)}. In these PDI images the suspected disk cavity inside of $\sim$15 au \citep{Bouwman et al.(2003),Grady et al.(2005)} was already tentatively detected and subsequent PDI data with higher signal-to-noise clearly confirmed a drop of polarized flux inside of these separations \citep{Avenhaus et al.(2014b)}. Those data also revealed a brightness asymmetry in the inner $\sim$0.5$''$ along the disk major axis (see, Figure~\ref{fig:PDI}), which is difficult to explain as an effect from the scattering or polarization phase function if one assumes similar dust properties on both sides of the star. Direct and indirect evidence for a companion orbiting inside the disk cavity are manifold \citep[e.g.,][]{Bouwman et al.(2003),Acke & van den Ancker(2006),Mulders et 
al.(2013)}, but the most compelling empirical result suggesting the existence of such a companion is probably the change in the spectro-astrometric signal of the CO v = 1-0 line over a time baseline of 10 years consistent with orbital motion \citep[][]{Brittain et al.(2013),Brittain et al.(2014)}. The line flux is interpreted as coming from a warm ($\sim$1400 K), $\sim$0.1 au$^2$ sized circumplanetary disk surrounding a young planetary companion \citep{Brittain et al.(2014)}. 

Motivated by the large scale spiral-arm structure of the disk initially seen in \emph{HST} observations\footnote{Additional spiral arm features closer to the star have recently been reported in \citet{Boccaletti et 
al.(2013)}, \citet{Avenhaus et al.(2014b)} and \citep{Currie et al.(2014)}.} \citep{Grady et al.(2001),Augereau et al.(2001),Ardila et al.(2007)}, high-contrast ADI observations in the L$'$ filter were carried out and resulted in the surprising discovery of another companion candidate $\sim$0.46-0.48$''$ away from the central star \citep{Quanz et al.(2013a)}. This object is located right in the middle of the circumstellar disk, where the PDI images did not show any peculiar morphological structure \citep{Avenhaus et al.(2014b)}\footnote{\citet{Avenhaus et al.(2014b)} showed that the disk 'wedge' initially detected in earlier PDI data \citep{Quanz et al.(2011),Quanz et al.(2013a)} was a calibration artifact.}. The young planet candidate was meanwhile re-detected in new L$'$ ADI data by \citet{Currie et al.(2014)}, and common proper motion and a first multi-color analysis (detection in L$'$ and M$'$ and an upper limit in K$_s$) is presented in \citet{Quanz et al.(2014)}. The observed properties of the source are best explained with a young forming gas giant planet that is likely surrounded by a circumplanetary disk, and recent ALMA observations that spatially resolved the mm-sized dust grains of the circumstellar disk provided hints for dynamical interactions between the disk and the forming planet \citep{Pineda et al.(2014),Walsh et al.(2014)}. 

Similar to HD169142 also HD100546 may harbor two young planets orbiting within its disk, but in the context of PDI studies it is interesting to emphasize that - at the moment - the outer planet does not seem to leave a detectable imprint on the disk surface. This either means the planet is very young and/or not very massive or the local disk structure prevents the formation of a significant gap detectable by the PDI observations.

\section{Conclusions}
PDI is a direct imaging technique that gives access to the inner few tens of au around nearby protoplanetary disks and is hence very well suited to probe the formation sites of (some) gas giant planets. In recent years, PDI on 8-m class telescopes with updated calibration procedures provided some breakthrough results by resolving a variety of distinct sub-structures (e.g., cavities, holes, spiral arms) in a number of protoplanetary disks around Herbig Ae/Be stars. Even though these disks were imaged previously in scattered light, most of these sub-structures were not detected until the PDI studies were carried out. 

Whether or not -- or to what extent -- the detected structures in Herbig Ae/Be disks are directly linked to ongoing or recent planet formation is a matter of active research. While the first candidate for a young planet orbiting within the gap of a transition disk was detected around a TTauri star \cite[LkCa 15,][]{Kraus 
& Ireland(2012)}, there is strong -- and further growing -- observational evidence that also some of the Herbig Ae/Be disks do host young planets. Currently the best examples are HD169142 and HD100546, where high-contrast imaging observations have revealed one young planet candidate around each star. Other observational techniques provided evidence for a second companion in each system and hence HD169142 and HD100546 are possibly the first two systems where the formation of multiple planets and their interaction with the protoplanetary disk can be studied empirically.
 
It is worth re-emphasizing that in both cases, where planetary companion candidates have been directly imaged inside the circumstellar disk of their host stars, they appear overluminous in the L$'$ band, i.e., between 3--4 $\mu$m. According to model predictions and taking the L$'$ flux as reference point they also should have been detected at shorter wavelengths, but they were not \citep{Reggiani et al.(2014),Biller et al.(2014)}. An elegant way out is to invoke the existence of a circumplanetary disk that has a larger emitting area but a lower effective temperature than the young planet itself. First model predictions for the brightness of circumplanetary disks do indeed predict high fluxes longwards of 3 $\mu$m \citep{Zhu(2015),Eisner(2015)}. 

The past few years  clearly demonstrated that PDI can be used to identify excellent targets for planet formation studies and dedicated follow-up observations. The objects with young planet candidates seem to indicate that gas giant planets may well form a few tens of au away from their central stars, at least around intermediate mass stars. This result is in particular interesting because it directly links to the imaged planets on very long-period orbits listed in Table~\ref{table1}: These objects may indeed have formed in situ or close to their current location. The underlying physical processes leading to the formation of very long-period gas giant planets is still to be investigated, but the Herbig Ae/Be stars discussed above offer a unique opportunity to address this question. 

\section{Future prospects}
The future prospects for high-contrast imaging planet formation studies are bright. Recently, new, dedicated high-contrast imagers came online: GPI at Gemini \citep{Macintosh et al.(2006)} and SPHERE at the VLT \citep{Beuzit et al.(2006)}. Both instruments are equipped with high-performing AO-systems providing unprecedented Strehl ratios at optical and NIR wavelengths from the ground and both instruments have PDI capabilities. The hope is that these instruments will help to significantly increase the number of protoplanetary disks with clearly resolved sub-structures and hence to provide additional high-profile targets for searching young, forming gas giant planets. As a matter of fact, as this article was in the writing, first preliminary PDI results from commissioning and science verification runs of these instruments started to become available and they showed the great promise of PDI for the coming years \citep[e.g.,][]{Rodigas et al.(2014)}. Of particular interest is ZIMPOL, the optical imaging polarimeter that is a sub-system of SPHERE \citep{Schmid et al.(2006b),Beuzit et al.(2008)}, as it offers a $\sim$2--3 times higher spatial resolution than the NIR images presented here. Hence, finer disk structures can be resolved and the inner working angle can be further decreased. 

\begin{figure}[t]
\centering
\includegraphics[height=4cm]{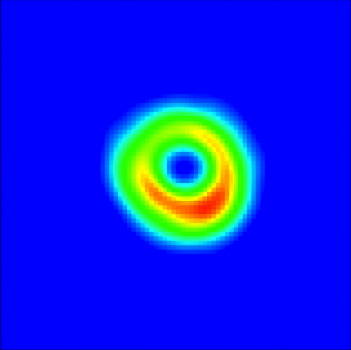}
\includegraphics[height=4cm]{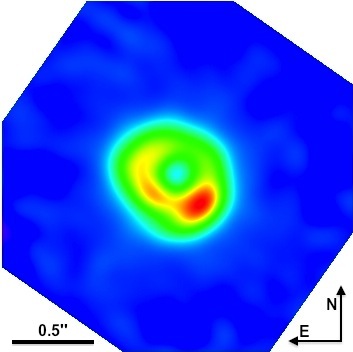}
\caption{Comparison between an ALMA 450 $\mu$m Cycle 0 image \citep[left, data from][]{Perez et al.(2014)} and a VLT/NACO scattered light image \citep[right,][]{Garufi et al.(2013)} of SAO206462. Both images are shown on the same spatial scale and are normalized to their peak fluxes. The NACO image has been convolved with the ALMA beam to have the same spatial resolution.
} %% no full stop at the end of caption
 \label{fig:SAO}
 \end{figure}

At the same time, GPI and SPHERE will help constrain the occurrence rate of gas giant planets on long-period orbits. The expectations are that these instruments will push the direct imaging detection limits for gas giant planets significantly closer to the central stars (on average inside of 20--30 au or so). These results will be of utmost importance to understand whether the systems listed in Table~\ref{table1} are indeed exceptions or whether there is a detectable increase in the occurrence rate of massive planets as one probes closer and closer to the star. Furthermore, depending on the final statistics of the surveyed stars, maybe one will be able to say whether the trend observed in radial velocity studies, namely that more massive stars have more massive planets further out, is robustly confirmed out to a few tens of au.

In addition to simply imaging more gas giant planets, GPI and SPHERE will also be able to characterize the detected planets to a great extent by sampling their SEDs with low resolution spectroscopy from the red optical wavelength range up to the K band. These data will constrain the composition of the planets' atmospheres \citep[e.g.,][]{Konopacky et al.(2013)}, which may in turn provide crucial information about the formation environment of these objects \citep{Oberg et al.(2011)}. 

Last, but certainly not least, a great leap forward in our understanding of gas giant planet formation can be expected from combining the results from (future) high-contrast imaging studies with observations at longer wavelengths, but with similar resolution (e.g., ALMA, EVLA, SKA). In particular ALMA, as of cycle 3, will provide a spatial resolution of $<$0.1$''$ in the (sub)-mm regime and probe circumstellar disk regions and components that are perfectly complementary to those probed by PDI. In particular, ALMA is sensitive to the population of larger, mm-sized dust grains in the cooler outer regions of circumstellar disks and in the disk mid-plane, and ALMA can detect the gaseous disk component by observing the emission from various molecules. Taking all this together one will be able to produce 3D pictures of numerous disks in gas and dust and derive physical and chemical parameters spatially resolved across these disks. One will be able to test whether the structures seen  in PDI images have counterparts at longer wavelengths and how discrepancies and similarities may be related to planet formation activities. As seen in section 5, there are some Herbig Ae/Be stars where already existing (sub)-mm data in combination with existing PDI data suggest the presence of planet-disk interactions and increased spatial resolution and sensitivity will offer additional insights and open up new discovery space. Figure~\ref{fig:SAO} illustrates this aspect further: Comparing a PDI image of SAO206462 with an early ALMA image, where the PDI data have been convolved with the ALMA beam to mimic the same spatial resolution, reveals that the overall morphology is not too different. A perfect match was not to be expected as different physical processes are responsible for the detected emission, and yet there is clear resemblance between the two images. Turning this around and looking again at the original PDI image of SAO206462 in Figure~\ref{fig:PDI} suggests that ALMA might also start resolving spiral arms in this disk once it has comparable spatial resolution. Indeed, \citet{Perez et al.(2014)} found first hints in this direction as the residuals, once a model consisting of a dust ring and a vortex-like structure was subtracted from the ALMA data, were suggestive of spiral arm features. That underlines once again the great prospects for ALMA in the coming months and years. 

Finally, with the high-spatial resolution achievable as of Cycle 3, ALMA will be able to probe for emission from the suspected circumplanetary disks surrounding some of the directly imaged young, planet candidates discussed above. There is a chance that one might be able to constrain the masses and fundamental physical parameters (e.g., temperatures) of these disks and maybe even get an idea about their gaseous composition. Such observations, in combination with information about the local circumstellar disk properties, will eventually empirically constrain the physical processes involved in the formation of gas giant planets on long-period orbits, and Herbig Ae/Be stars offer a unique opportunity that should not be missed. 

%% Math 
%
%\begin{eqnarray}%\label{eqn:?}
%\\ \nonumber
%\end{eqnarray}
%
%\begin{equation}%\label{eqn:?}
%\end{equation}

%% Table (one-colum)
%
% \begin{table}
% \caption{} %% no full stop at the end of caption
% \label{tbl:?}
% \begin{tabular}{}
% \tableline  %% rule at top
% \tablenotemark{a} 
% <entries>
% \tableline %% rule at bottom
% \end{tabular}
% \end{table}

%% Deluxe tabel (refer to AASTeX documentation)
%

%% Acknowledgements
%
\acknowledgments
I would like to thank the referee for a constructive report and H. Avenhaus, A. Garufi, and H.M. Schmid for helpful comments and suggestions that helped improve the quality of the manuscript.  Furthermore, I am grateful to the editors of the Topical Collection on HAeBe stars (Willem-Jan de Wit and Rene Oudmaijer) for inviting me to contribute this work and I thank all colleagues who gave permission to reproduce figures from their original publications. 
Part of this work has been carried out within the frame of the National Centre for Competence in Research PlanetS supported by the Swiss National Science Foundation. The author acknowledges the financial support of the SNSF.

%% References
%% Please cite all reference entries in the article text using \cite or
%% equivalent command. 

%%%  Using BibTeX  (Name-Year style)
%
% \bibliographystyle{spr-mp-nameyear-cnd}  %% BibTeX style
% \bibliography{<bib data>}                %% BibTeX data

%% Non-BibTeX  (Name-Year style)
%

\end{document}